\documentstyle{mn}
\input{epsf.sty}
\newcommand{\figpath}{.}
\def\plottwo#1#2{\centering \leavevmode
\epsfxsize=.95\columnwidth \epsfbox{#1} \hfil
\epsfxsize=.95\columnwidth \epsfbox{#2}}
\begin{document}

\title[Coplanar disc-disc encounters]{Numerical simulations of protostellar encounters \\ II. Coplanar disc-disc encounters}

\author[S.J.~Watkins et al.]
  {S.J.~Watkins,
  A.S.~Bhattal,
  H.M.J.~Boffin,
  N.~Francis
  and A.P.~Whitworth\\
  Department of Physics and Astronomy, University of Wales, Cardiff CF2 3YB, Wales, UK.
  }

\maketitle
\begin{abstract}

It is expected that an average protostar will undergo at least one impulsive interaction with a neighbouring protostar whilst a large fraction of its mass is still in a massive, extended disc. Such interactions must have a significant impact upon the evolution of the protostars and their discs.

We have carried out a series of simulations of coplanar encounters between two stars, each possessing a massive circumstellar disc, using an SPH code that models gravitational, hydrodynamic and viscous forces. We find that during a coplanar encounter, disc material is swept up into a shock layer between the two interacting stars, and the layer then fragments to produce new protostellar condensations. The truncated remains of the discs may subsequently fragment; and the outer regions of the discs may be thrown off to form circumbinary disc-like structures around the stars. Thus coplanar disc-disc encounters lead efficiently to the formation of multiple star systems and small-${\cal N}$ clusters.

\end{abstract}
\begin{keywords}
stars: formation -- binaries: general -- accretion, accretion discs -- methods: numerical -- hydrodynamics -- instabilities
\end{keywords}

\section{Introduction}
 
Most main sequence stars are observed to be in binaries or higher order groupings (Duquennoy \& Mayor 1991\nocite{duquennoy:mayor}, Abt \& Levy 1976\nocite{abt:levy}, Fischer \& Marcy 1992\nocite{fischer}). For instance, Duquennoy \& Mayor \shortcite{duquennoy:mayor} find a binary frequency of 60\% for main sequence solar-type stars (the binary frequency is defined as the percentage of stars that are in multiple systems, as opposed to the binary fraction, which is the percentage of systems that are multiple).

Much work has been done recently on detecting binary companions of T-Tauri stars. Observations have been carried out in both Taurus-Auriga and Scorpius-Ophiuchus, using a number of complementary techniques, such as spectroscopy \cite{mathieu89}, speckle imaging (Ghez, Neugebauer \& Matthews 1993\nocite{ghez}, Leinert et al.\ 1993\nocite{leinert}), solid state arrays \cite{leinert} and lunar occultation \cite{simon}, which together can detect binary companions with projected separations in the range 3-1400 AU (at a distance $\sim$ 150 pc). They find a binary frequency that is comparable to, or greater than, that observed for solar-type main-sequence stars.
Binary and multiple systems are therefore formed before the T-Tauri stage, while protostars still possess massive, extended discs.
 
Most stars are born in clusters \cite{lada91}, in which the separations between stars are comparable to the sizes of the discs observed around young stars (Lada et al.\ 1991\nocite{lada91}, Strom et al.\ 1989\nocite{strom89b}). Consequently, most stars are likely to undergo at least one encounter with a neighbouring protostar over the lifetime of the protostellar disc, and, given the apparent coevality of star formation (e.g.\ Gauvin \& Strom 1992\nocite{gauvin:strom}), a significant fraction of these encounters will be disc-disc rather than disc-star encounters.

Pringle \shortcite{pringle89} has suggested that the protostars formed from the fragmentation of a rotating molecular cloud will be surrounded by discs whose sizes are comparable to the periastron separations of the protostars. In this situation, interactions between the discs are an inevitable consequence of the formation process.

Disc-disc interactions are likely to play a vital r\^{o}le in more dynamic star formation as well, for instance when star formation is triggered by the collision of two clumps within a molecular cloud. Numerical simulations (Chapman et al.\ 1992\nocite{chapman92}, Turner et al.\ 1995\nocite{turner}, Whitworth et al.\ 1995\nocite{whitworth}) indicate that such collisions lead to the formation of a shock layer that becomes gravitationally unstable and fragments, collapsing to form a number of massive protostellar discs. Two of these discs may then fall together and undergo a strong tidal interaction, resulting in the formation of a binary system.

It therefore appears that interactions between protostellar discs may play an important part in determining the properties of the resulting stars, both in quiescent star-formation (where two separately-born protostars drift together and interact) and in highly-dynamic star formation (e.g.\ cloud-cloud collisions).
 
We have carried out a series of simulations of such disc-disc interactions.
This paper is the second of three describing simulations of protostellar encounters. Paper I \cite{watkins97a} described our numerical method, and presented the results of simulations of interactions between a star possessing an extended disc and a discless star. In this paper, we present the results of a series of simulations of coplanar encounters between two stars possessing extended, massive discs. Paper III \cite{watkins97c} presents corresponding results for non-coplanar disc-disc encounters, as well as an overall analysis of the results of the simulations of disc-disc encounters, and of the differences between such interactions and the disc-star interactions of Paper I.

\begin{table}
\begin{tabular}{l|l}
property & value   \\ \hline
disc radius & 1000AU \\
disc mass & 0.5$\mbox{M}_{\sun}$ \\
star mass & 0.5$\mbox{M}_{\sun}$ \\
disc density profile & $r^{-3/2}$ \\
eccentricity of orbit & 1.0 \\
initial separation & 5000AU \\
shear viscosity & $\nu_{s} = c_{s} h /800$ \\
\end{tabular}
\caption[Physical parameters used in simulations]
{\label{table:phys}Physical parameters used in simulations}
\end{table}

\section{Physical and computational model}

The numerical method used is described in Paper I; we also discuss its limitations in Paper I. The discs modelled in the simulations described in this paper have the same physical properties as the discs used in Paper I. A brief summary of these properties is given in Table \ref{table:phys}. Each disc was modelled with 2000 SPH particles -- so one particle represents a mass of $2.5 \times 10^{-4} M_{\odot}$.

There are three different types of coplanar disc-disc encounter, as opposed to the two types of coplanar disc-star encounter. These are, adopting the notation used by Lattanzio \& Henriksen \shortcite{lattanzio:henriksen} for cloud-cloud collisions, spin-orbit parallel (SOP), in which the spins of both discs and the orbit are aligned, spin-orbit mixed (SOM), in which the spin of one of the discs is in the opposite sense to that of the other disc and the orbit, and spin-orbit anti-parallel (SOA), in which the spins of the two discs are in the same direction, and opposite to that of the orbit. The coplanar disc-disc encounters were carried out at the same four periastra as the disc-star encounters. The simulations are listed in Table \ref{table:runs}. In total, 12 simulations of coplanar disc-disc interactions were carried out.

The simulations are labelled so that the number of the disc-disc simulation is the same as that of the corresponding disc-star encounter from Paper I. In the SOP simulations (dd01-dd04), the spin of the orbit is in the same direction as the discs, and so is similar to the coplanar prograde disc-star encounters ds01-ds04. The SOA simulations (dd21-dd24) have the orbital spin opposed to the disc spin for both discs, whilst the SOM encounters (dd17-dd20) have the orbital spin opposed to the spin of one disc, but parallel to the other. Both of these can be compared with the retrograde disc-star encounters (ds17-ds20) in Paper I.

\begin{table}
\begin{tabular}{|c|c|c|c|c|r|r|} \hline \hline
Run    & $r_{peri}$ & type   \\ \hline \hline
dd01-04 & 500, 1000, 1500, 2000 & SOP \\ \hline
dd17-20 & 500, 1000, 1500, 2000 & SOM \\ \hline
dd21-24 & 500, 1000, 1500, 2000 & SOA \\ \hline
\end{tabular}
\caption[]{\label{table:runs}
List of simulations}
\end{table}

\section{Simulations}

The figures presented in this section show the results of three of the simulations conducted. The figures are grey-scale plots of surface density, in which the shading is logarithmically-scaled. These plots are overlaid with contours of constant surface density, that are equally separated in log-space. Occasionally an extra contour has been added in order to highlight an important feature of the figure, such as a shock or spiral arm. The contour levels used are given in the caption to the figure. The time at each figure, in units of $10^{4}$ years, is also given, and is calculated from the start of the simulation, when the two stars are at a separation of 5$r_{disc}$. The individual frames within a figure, when referred to within the text or a caption, are labelled from top left to bottom right in rows, so that frames 1 and 2 are on the top row, frames 3 and 4 on the next row and so on.

\subsection{run dd01 : spin-orbit parallel}

\begin{figure*}
\plottwo{\figpath/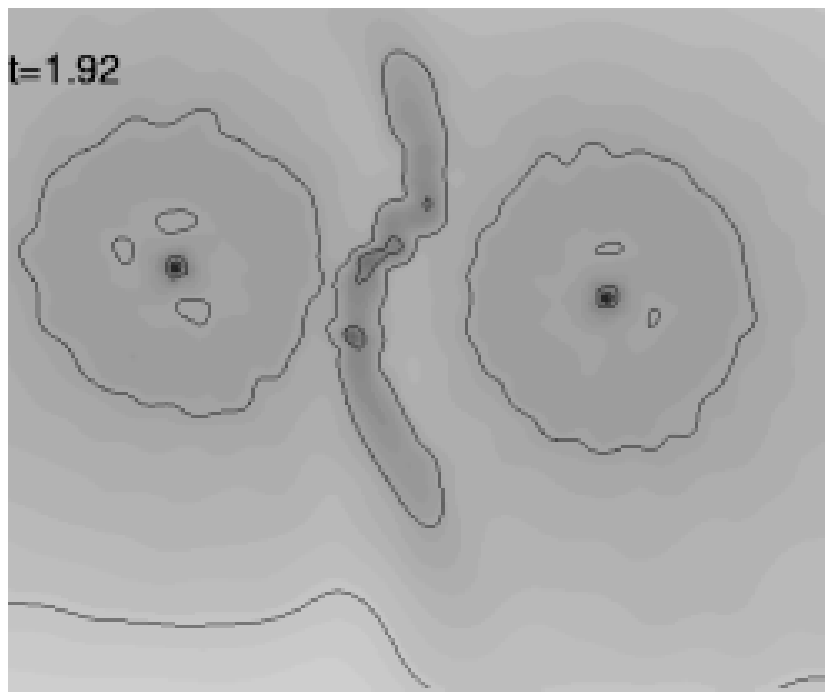}{\figpath/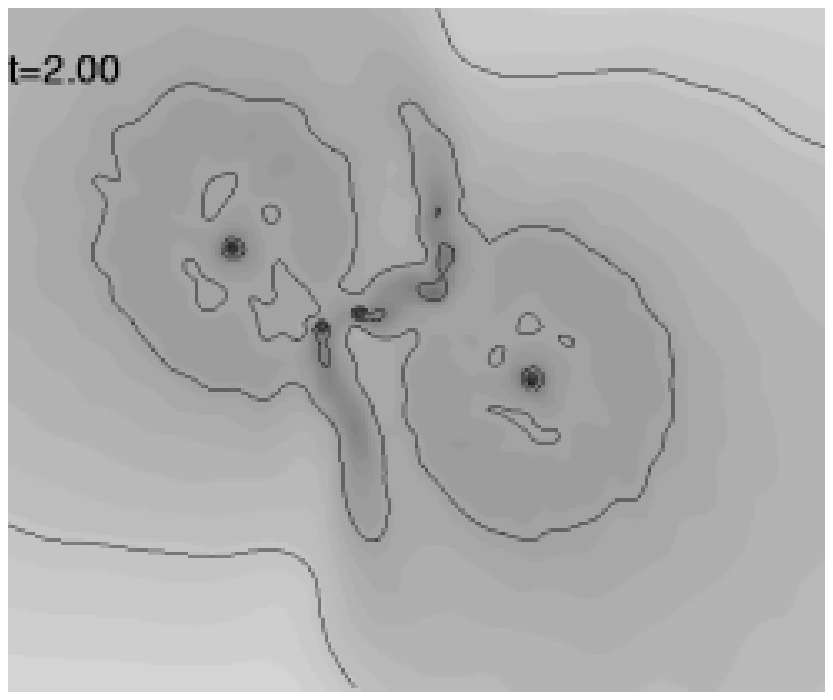}\\
\vspace{6pt}
\plottwo{\figpath/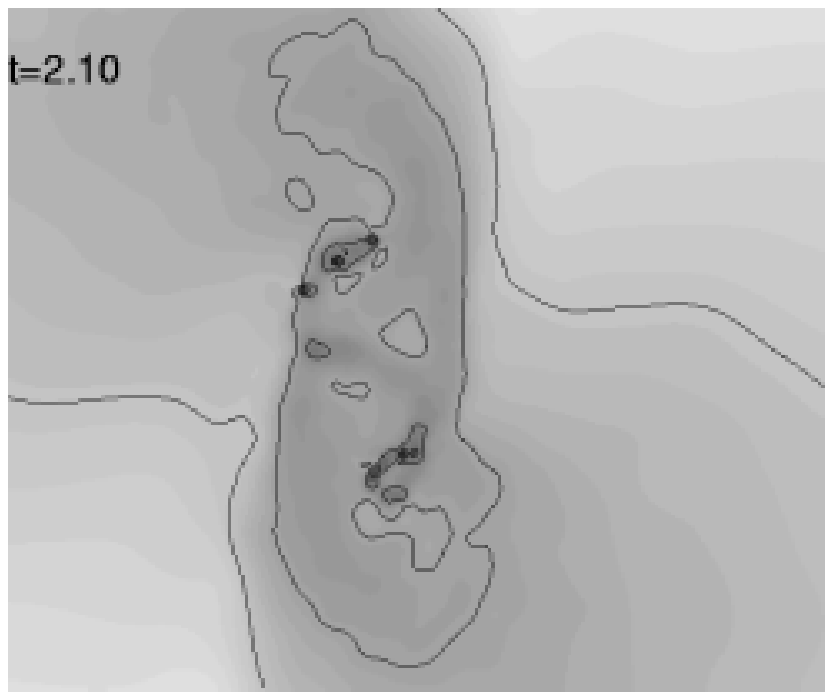}{\figpath/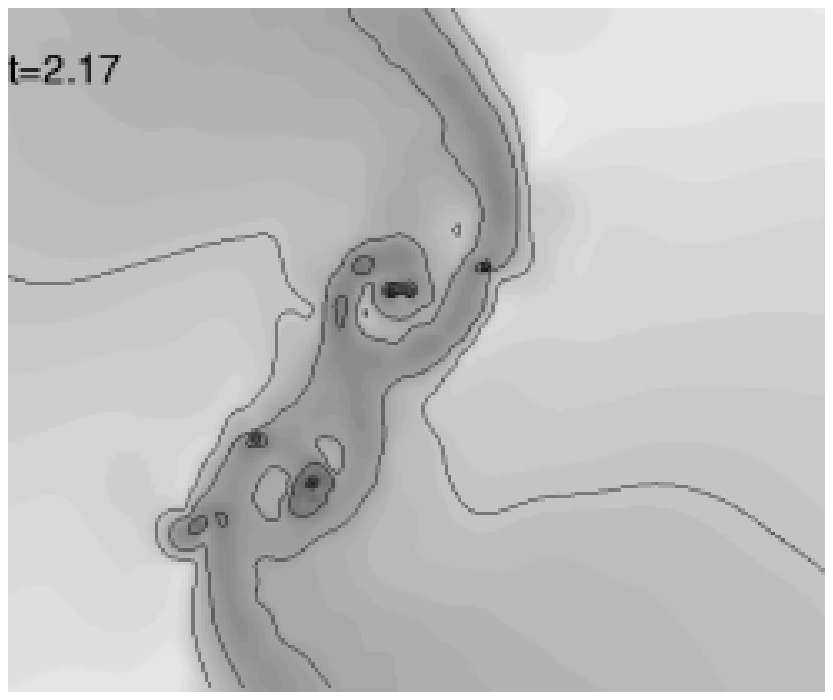}\\
\vspace{6pt}
\plottwo{\figpath/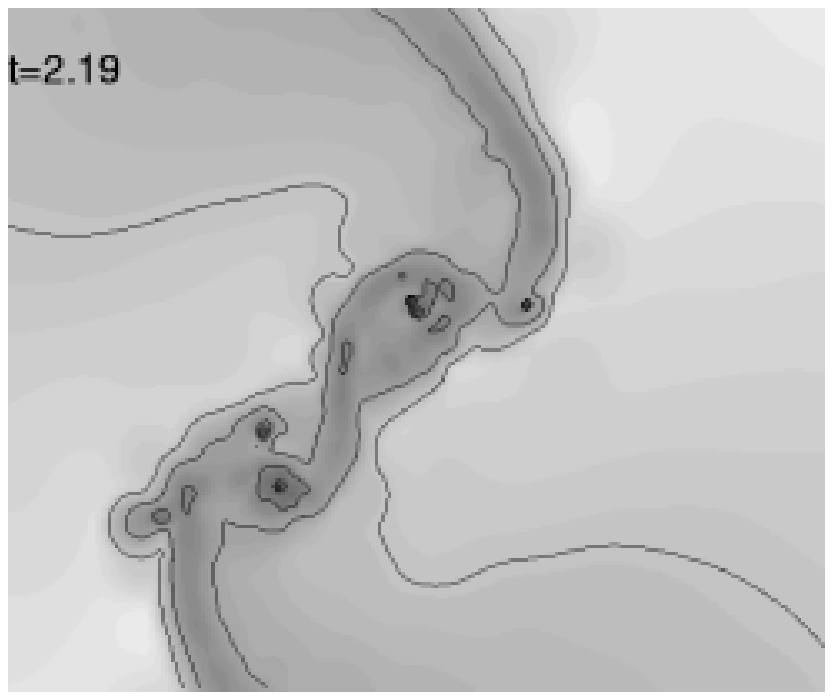}{\figpath/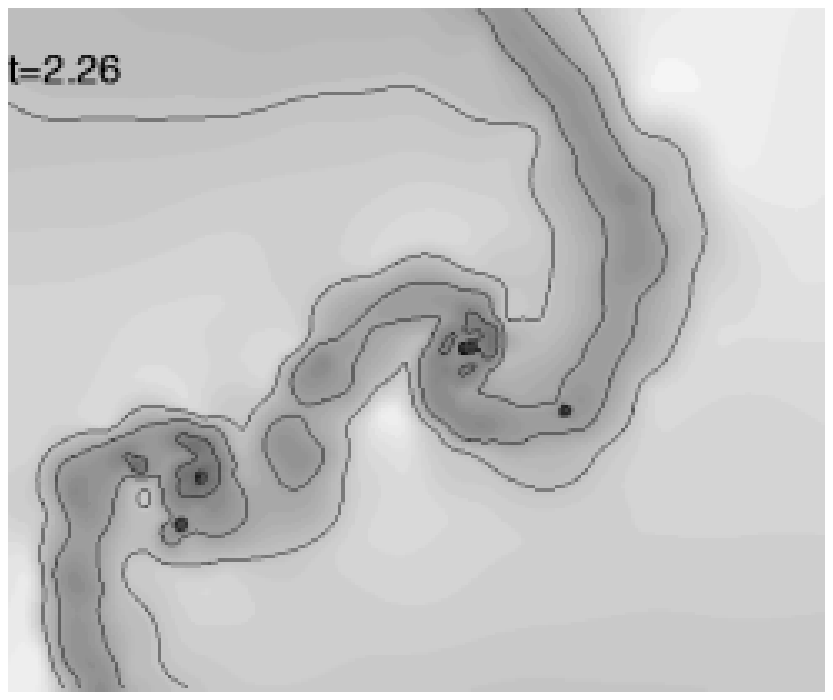}\\
\caption[Simulation dd01 : SOP, $r_{peri}=500 \mbox{AU}$. 2000 $\times$ 1500 AU region.]{\label{fig:dd01a}Simulation dd01 : SOP, $r_{peri}=500 \mbox{AU}$. 2000 $\times$ 1500 AU region. Contour levels at [0.3, 3, 30, 300] $\mbox{g cm}^{-2}$. The primary is initially on the left, and orbits clockwise about the secondary. As the two discs collide, a shock forms which buckles and fragments to produce new condensations, some of which are tidally disrupted, others of which survive as new protostars. The outer regions of the discs form long tidal tails. The end state is two binary systems.}
\end{figure*}

\begin{figure*}
\plottwo{\figpath/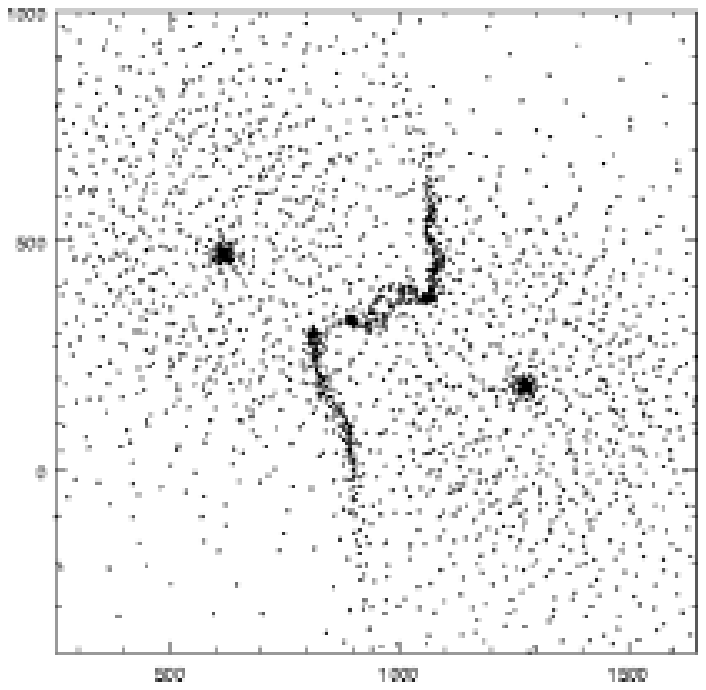}{\figpath/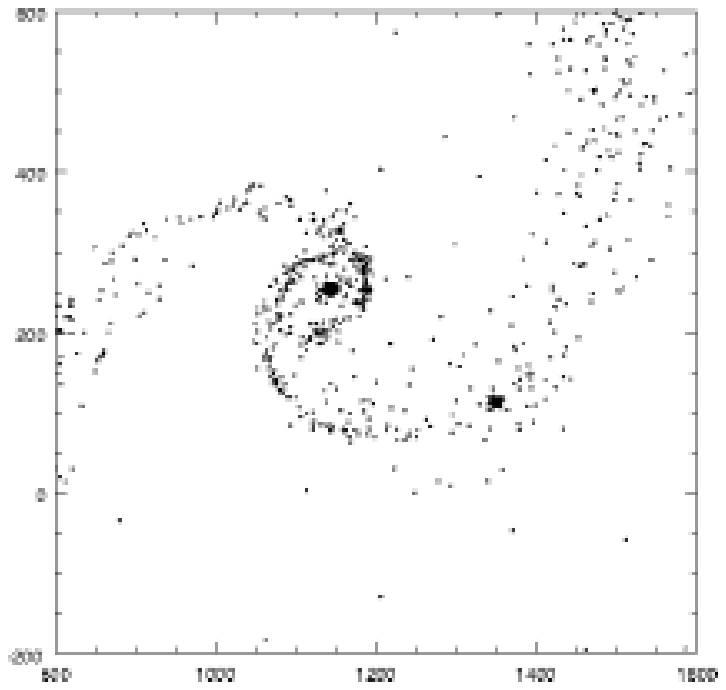}\\
\caption[Simulation dd01 - particle plot]{\label{fig:dd01c}Simulation dd01 : SOP, $r_{peri}=500 \mbox{AU}$. Positions of SPH particles. The left-hand figure shows the fragmentation of the shock layer, and corresponds to frame 2 of the previous figure. The right-hand figure shows the primary star and its newly-formed companion, at a time corresponding to frame 6 of the previous figure.}
\end{figure*}

Spin-orbit parallel encounters are those in which the spins of both discs and the orbit are parallel. Numerical simulations (Turner et al.\ 1995\nocite{turner}, Whitworth et al.\ 1995\nocite{whitworth}) suggest that during star formation triggered by collisions between clumps in molecular clouds, the majority of disc-disc interactions that occur may approximate to coplanar spin-orbit parallel encounters.

The results of run dd01 are shown in Fig.~\ref{fig:dd01a}. The orbit and disc spins are clockwise. The primary is initially on the left-hand side of the diagram, and moves towards the upper right side over the course of the encounter, whilst the secondary correspondingly moves towards the lower left. The periastron distance for the encounter is 500AU, compared with a disc radius of 1000AU.

As the discs collide, disc material between the stars becomes swept up into a shock layer. As the collision proceeds, the shock layer experiences a torque due to the fact that the collision is not head-on. This causes the layer to become bent, and also to rotate. Once enough mass has been swept up into the layer, it also becomes gravitationally unstable and starts to fragment. In this case the layer produces three fragments. One of the fragments is tidally disrupted by the secondary, whilst the other two fragments fall onto the primary star. One of these falls directly onto the primary star and merges with it, while the other is captured by the primary to form a binary system. Meanwhile the remains of the fragment that was disrupted by the secondary recondense and accrete from the secondary disc to form a binary companion to the secondary star.

At the same time, both of the discs undergo strong $m=2$ spiral arm instabilities, in which the inner arm connects the disc to the corresponding arm on the other disc, and the leading edge of the disc is swept up into the second, outer arm. As the two binaries move away from each other, the inner arm breaks off and becomes closely wrapped around the binary, while the outer arms surround the two binary systems in a circumbinary disc-like configuration. Figure \ref{fig:dd01c} shows particle plots of the fragmentation of the shock layer, and of the primary star and its binary companion at the end of the simulation.

The primary star has captured 0.15$\mbox{M}_{\sun}$ into a small, unresolved disc. Its binary companion is also of mass 0.15$\mbox{M}_{\sun}$, and the two are in an orbit of eccentricity 0.4 and periastron 60AU. The secondary star has captured a disc of mass 0.15$\mbox{M}_{\sun}$ and radius 65AU. Its companion is of mass 0.08$\mbox{M}_{\sun}$, and has no circumstellar disc. These two stars are bound with an eccentricity of 0.2 and periastron 85AU. The two binary systems are marginally bound to each other on an orbit with eccentricity 0.99 and periastron 430AU.

\begin{figure*}
\plottwo{\figpath/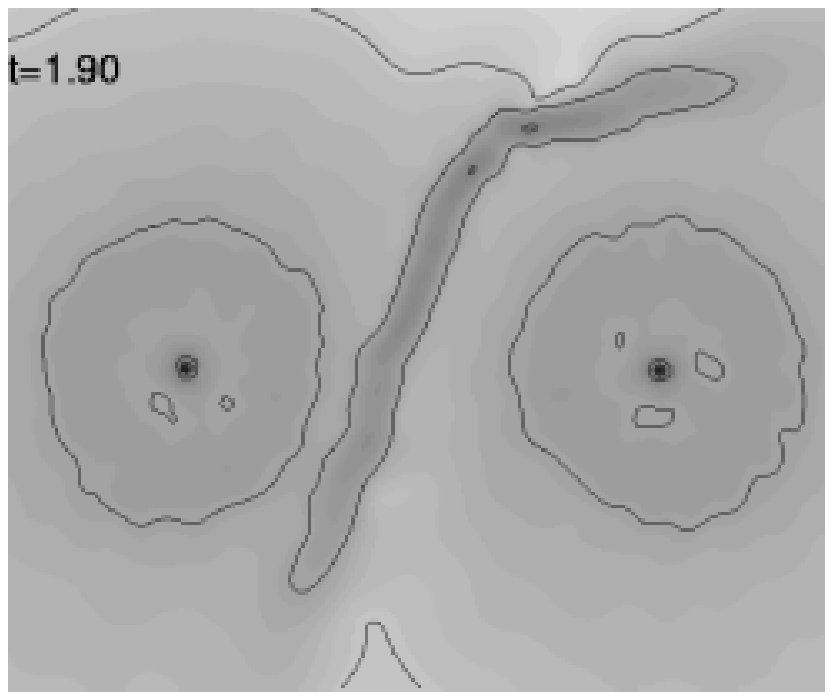}{\figpath/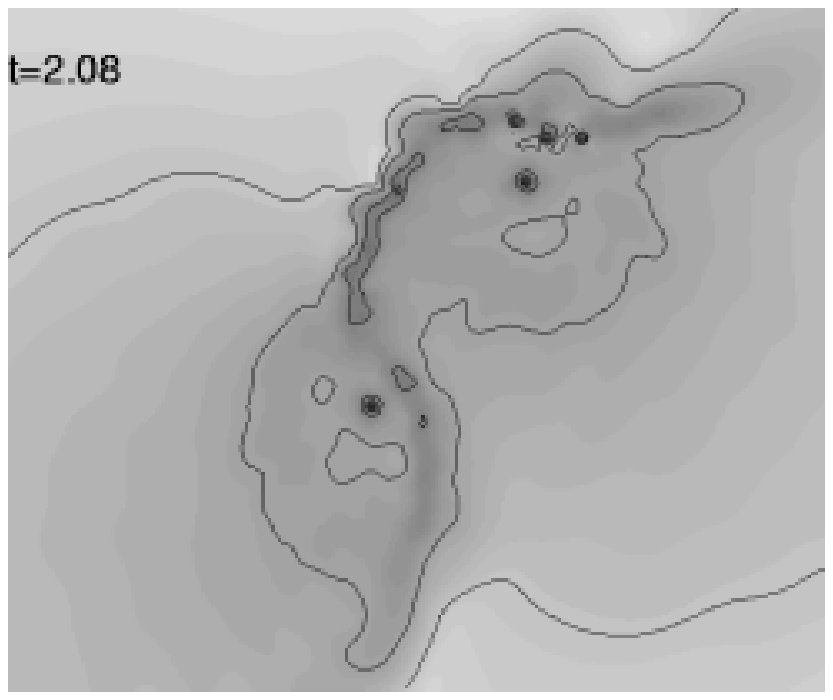}\\
\vspace{6pt}
\plottwo{\figpath/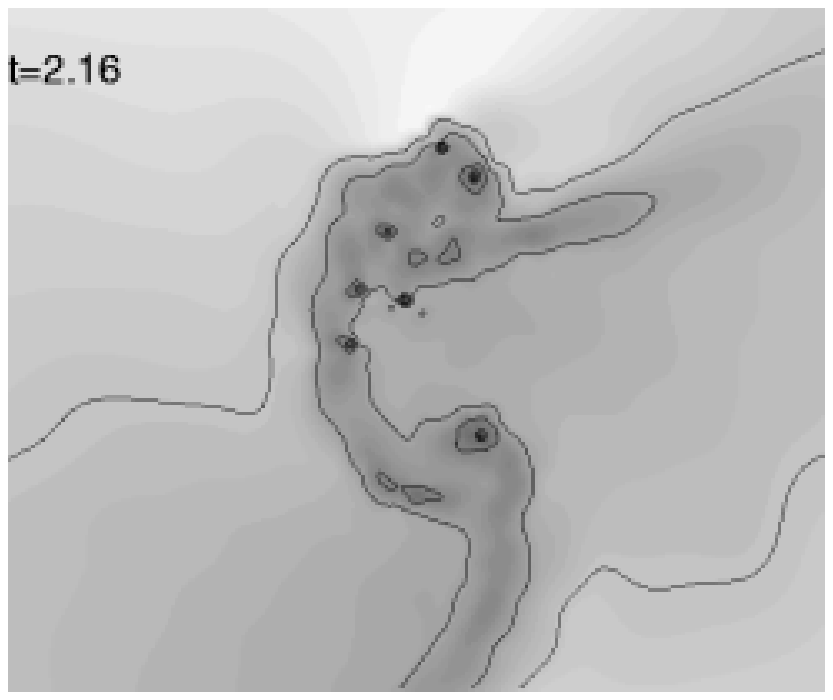}{\figpath/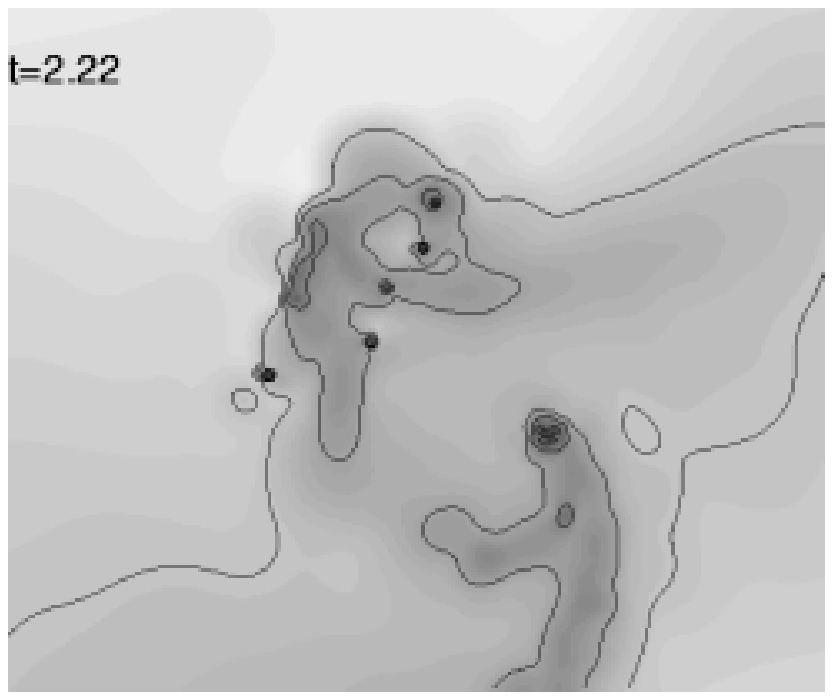}\\
\vspace{6pt}
\plottwo{\figpath/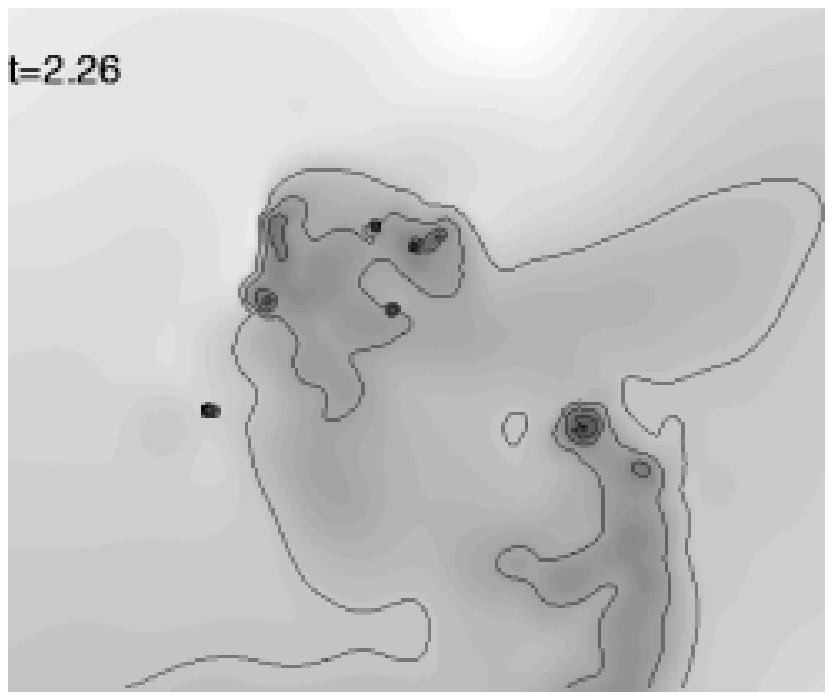}{\figpath/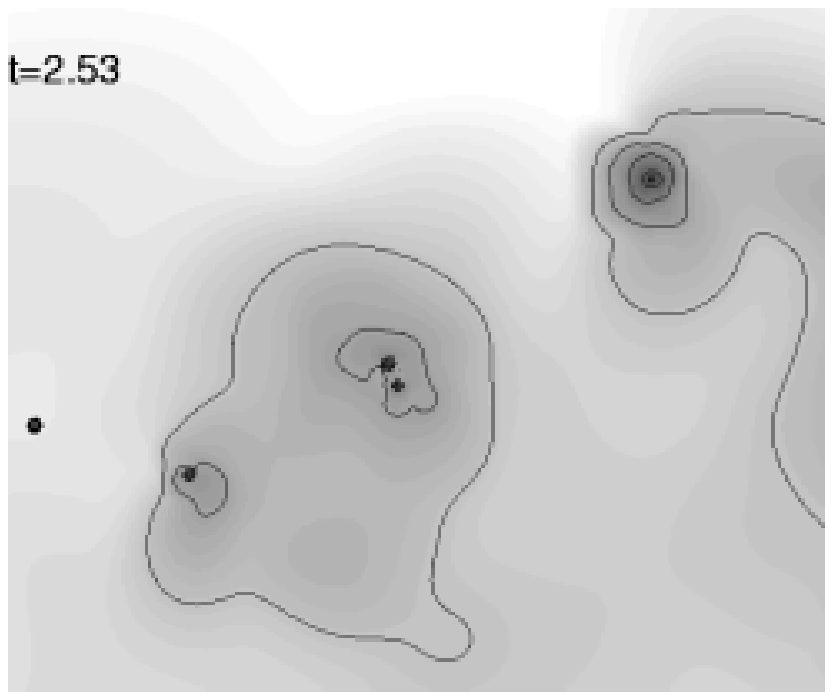}\\
\caption[Simulation dd17 : SOM, $r_{peri}=500 \mbox{AU}$. 2000 $\times$ 1500 AU region.]{\label{fig:dd17}Simulation dd17 : SOM, $r_{peri}=500 \mbox{AU}$. 2000 $\times$ 1500 AU region. Contour levels at [0.3, 3, 30, 300] $\mbox{g cm}^{-2}$. The primary disc, which is rotating clockwise, is initially on the right, and orbits anti-clockwise about the secondary disc, which is itself rotating anti-clockwise. The shock layer fragments to form a large number of condensations, some of which survive to form a bound cluster with the original stars.}
\end{figure*}

\begin{figure*}
\plottwo{\figpath/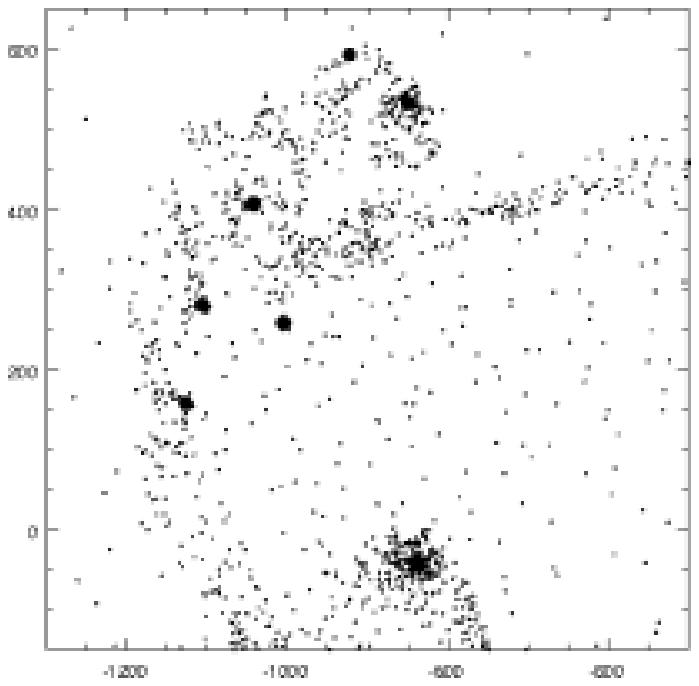}{\figpath/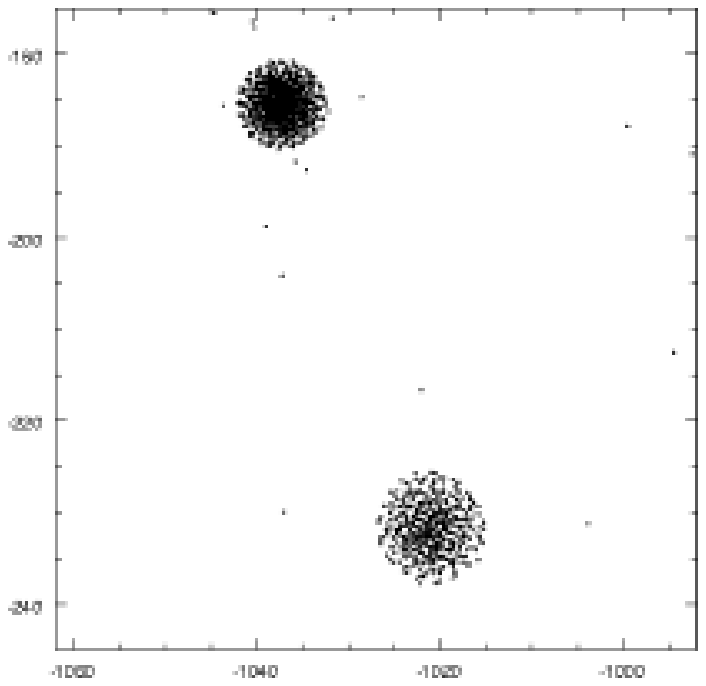}\\
\caption[Simulation dd17 - particle plot]{\label{fig:dd17b}Simulation dd17 : SOM, $r_{peri}=500 \mbox{AU}$. Positions of SPH particles. The left-hand figure shows the fragmentation of the shock layer, and corresponds to frame 3 of Fig.~\protect\ref{fig:dd17}. The right-hand figure shows the primary star and its newly-formed companion, at a time corresponding to frame 6 of Fig.~\protect\ref{fig:dd17}. The primary and its companion contain 962 and 531 particles respectively.}
\end{figure*}

\subsection{run dd17 : spin-orbit mixed}

In spin-orbit mixed (SOM) encounters, the spin of the secondary disc and orbit are anti-parallel to that of the primary disc. The mechanism that operates during these encounters is similar to that which operates during spin-orbit parallel (SOP) encounters, with the build-up of a shock layer that then fragments to produce new protostellar condensations, some of which are quickly disrupted, and others of which survive to form new multiple star systems.

This can be seen during run dd17 (Fig.~\ref{fig:dd17}), in which the periastron distance is 500AU. The primary disc is initially on the lower right-hand side of the figure and is spinning clockwise, whilst the secondary disc is on the upper left-hand side and is spinning anti-clockwise. As the two discs collide, material is swept up into a shock layer in the same way as for the SOP encounters. The shock is longer than in the SOP case, however, and produces a larger number of fragments, nine in this case, with all of the fragmentation occurring fairly close to the primary star. Two of the fragments merge, two are tidally disrupted by the primary star, and three fall directly onto the primary star and are accreted. Of the remaining three condensations, two experience slingshot events but remain bound to the system, and the third is captured by the primary star into a binary system. The secondary star, meanwhile, passes through the primary disc, accreting from it onto a small, massive circumstellar disc, while the secondary disc and the remainder of the primary disc are drawn out behind the secondary star in a long tidal tail.

The five stars are bound in a cluster. This is an unstable configuration, and there are likely to be a number of interactions and ejections before the system reaches a steady state. Of the two stars that have already experienced slingshot events, one is of mass 0.05$\mbox{M}_{\sun}$, and the other of mass 0.06$\mbox{M}_{\sun}$. The primary star has a small disc of mass 0.25$\mbox{M}_{\sun}$, whilst its binary companion is of mass 0.14$\mbox{M}_{\sun}$. The binary orbit has eccentricity 0.33 and periastron 33AU; this is tight enough that the binary may well survive any encounters that it undergoes with the other stars.
 
The secondary has a disc of mass 0.24$\mbox{M}_{\sun}$ and radius 80AU. It is bound to the cluster around the primary on an orbit with eccentricity 0.22 and periastron 530AU.

\begin{figure*}
\plottwo{\figpath/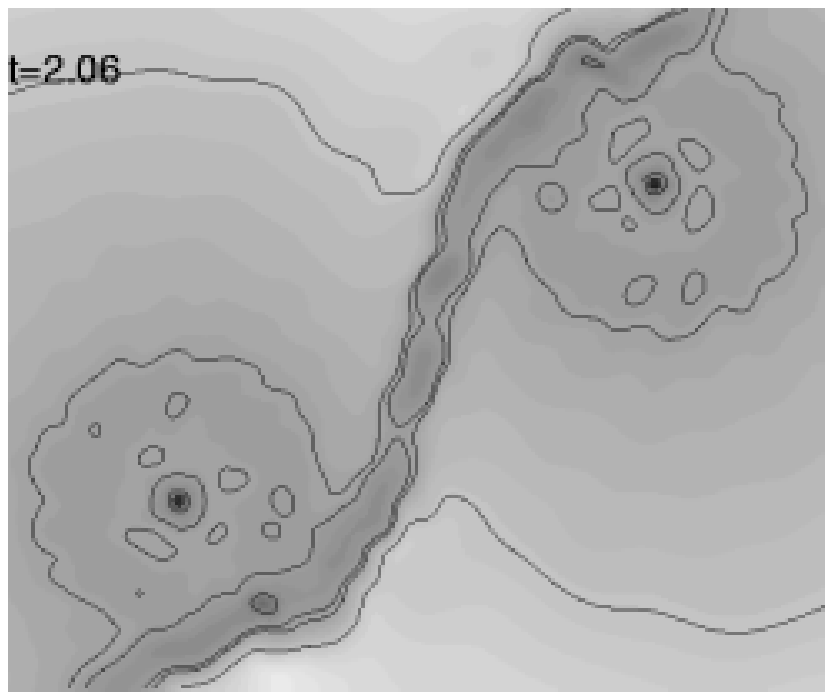}{\figpath/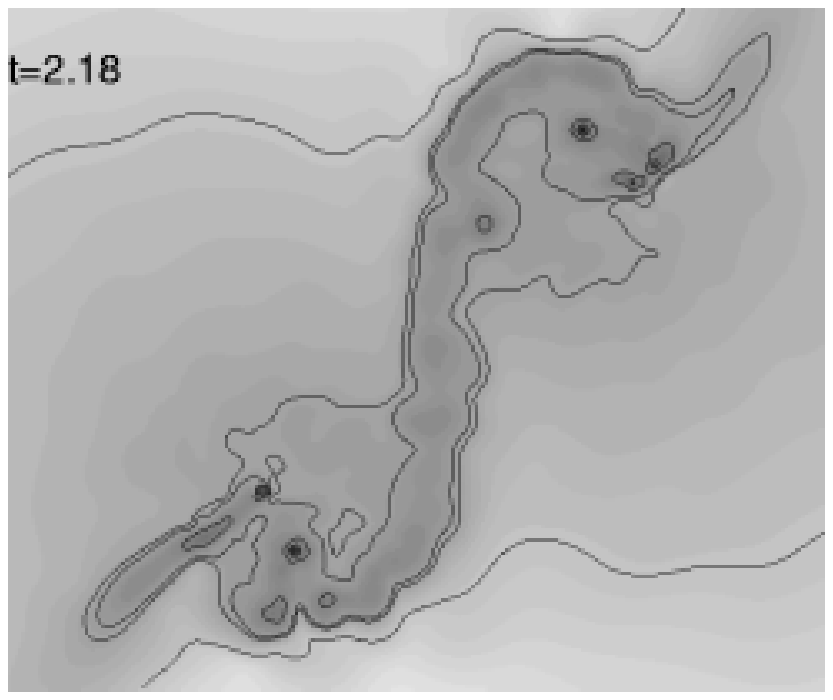}\\
\vspace{6pt}
\plottwo{\figpath/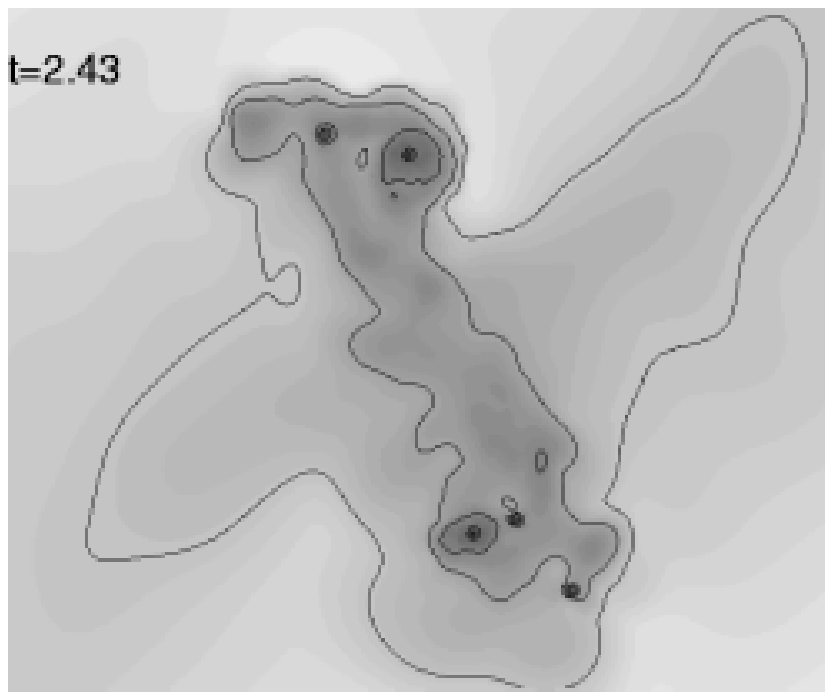}{\figpath/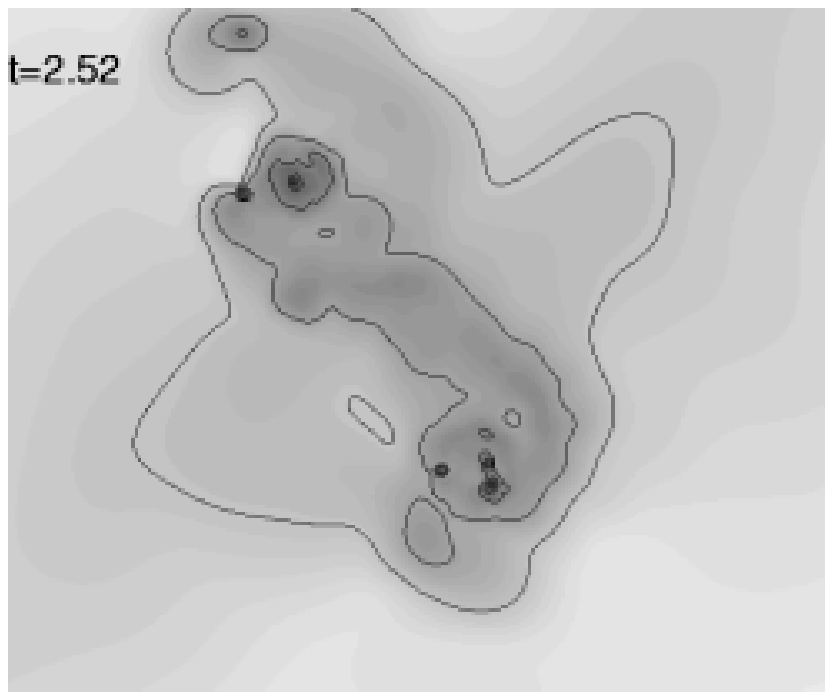}\\
\vspace{6pt}
\plottwo{\figpath/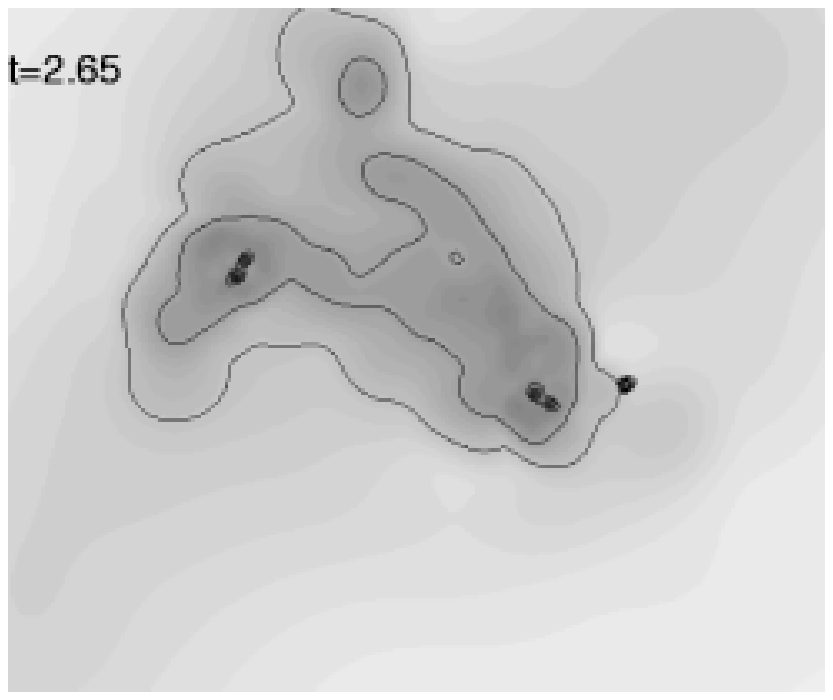}{\figpath/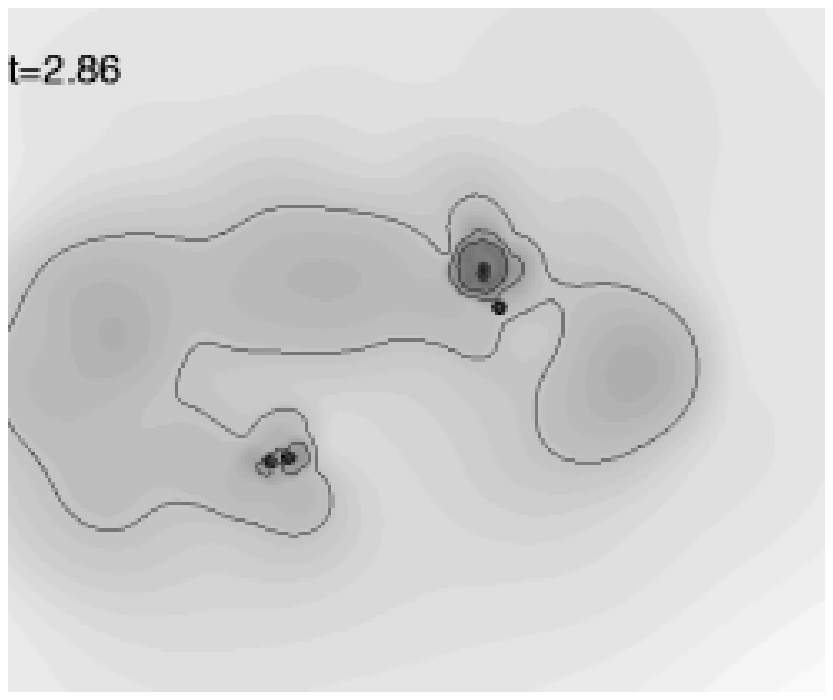}\\
\caption[Simulation dd22 : SOA, $r_{peri}=1000 \mbox{AU}$. 2000 $\times$ 1500 AU region.]{\label{fig:dd22}Simulation dd22 : SOA, $r_{peri}=1000 \mbox{AU}$. 2000 $\times$ 1500 AU region. Contour levels at [0.3, 3, 4 (frame 1, 2), 30, 300] $\mbox{g cm}^{-2}$. The two discs are rotating clockwise, while the orbit of the encounter is in an anti-clockwise sense. The primary disc is initially on the right-hand side. Fragmentation of the shock layer produces a binary and a triple system, before one of the condensations is disrupted leaving a hierarchical quadruple.}
\end{figure*}

\subsection{run dd22 : spin-orbit anti-parallel}

In spin-orbit anti-parallel (SOA) encounters, the spins of the two discs are aligned, and are in the opposite direction to that of the orbit.

Figure \ref{fig:dd22} shows the results of run dd22, in which the periastron is 1000AU. The mechanism that operates is the same as in the previous simulation, with disc material forming a shock layer that fragments at the ends to produce several new condensations. One of the fragments forms a binary with the primary star, while two more are captured by the secondary star to form a hierarchical triple. The triple system lasts for about ten orbital periods, before the inner condensation is tidally disrupted by the secondary star, leaving a binary system. We note parenthetically that, if gravity were not softenned, this condensation might have become sufficiently tightly bound to survive. While the triple system is in existence, the secondary star has accreted 0.15$\mbox{M}_{\sun}$, and its close companion is of mass 0.16$\mbox{M}_{\sun}$. These are on an orbit of eccentricity 0.16 and periastron 31AU. Orbiting them is the second condensation, which is of mass 0.04$\mbox{M}_{\sun}$, and on an orbit of eccentricity 0.41 and periastron 93AU. Once the inner star has been disrupted, the secondary is left with 0.36$\mbox{M}_{\sun}$ of accreted material, a 55AU radius disc of mass 0.08$\mbox{M}_{\sun}$, and a binary companion of mass 0.05$\mbox{M}_{\sun}$ on an orbit of eccentricity 0.35 and periastron 66AU.
The primary star, meanwhile, has accreted 0.21$\mbox{M}_{\sun}$ of disc material, and has a binary companion of mass 0.24$\mbox{M}_{\sun}$ on an orbit of eccentricity 0.14 and periastron 29AU. The two binaries are strongly bound to each other, with eccentricity 0.16 and periastron 520AU. The end state of the system is therefore an hierarchical quadruple.

\begin{figure*}
\plottwo{\figpath/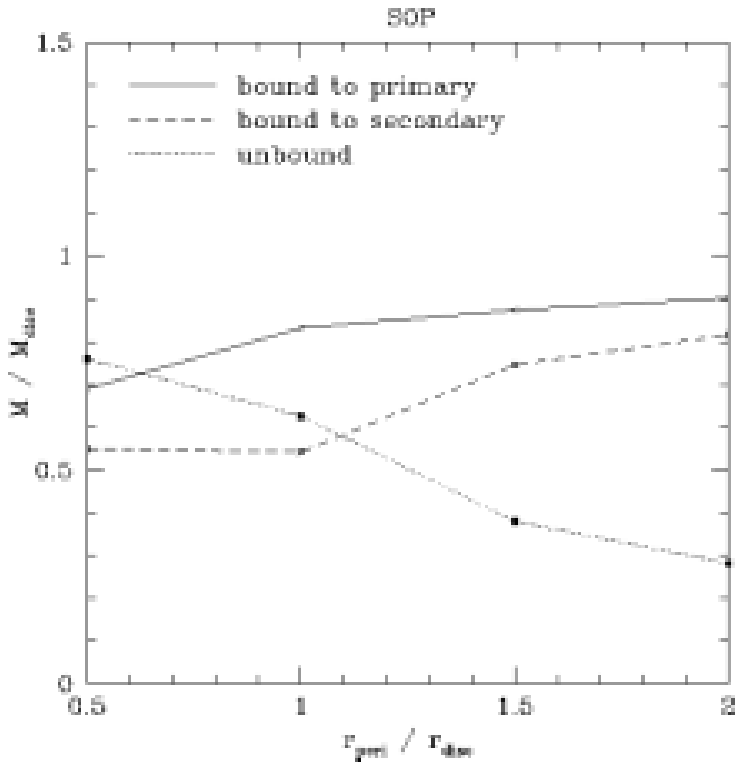}{\figpath/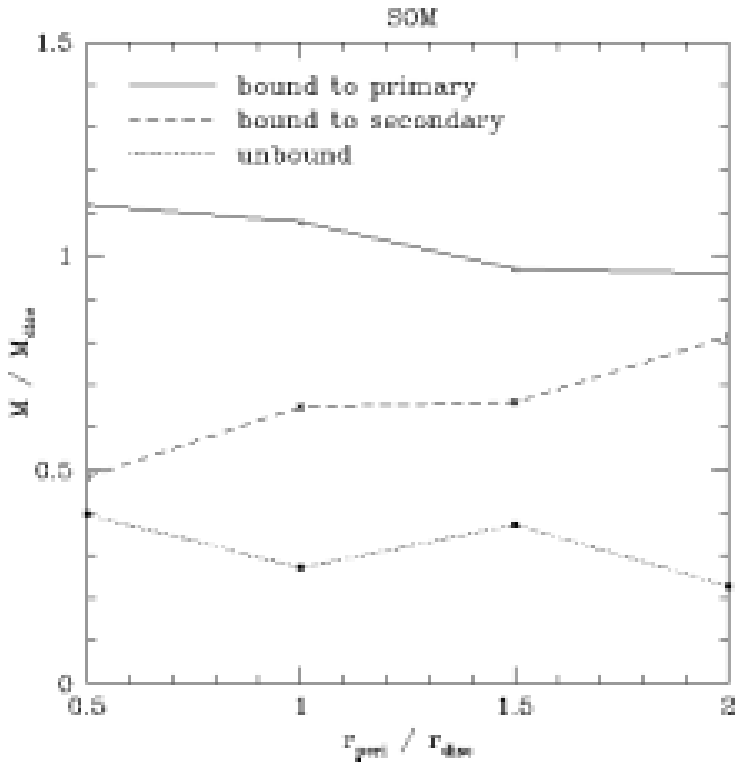}\\
\vspace{6pt}
\plottwo{\figpath/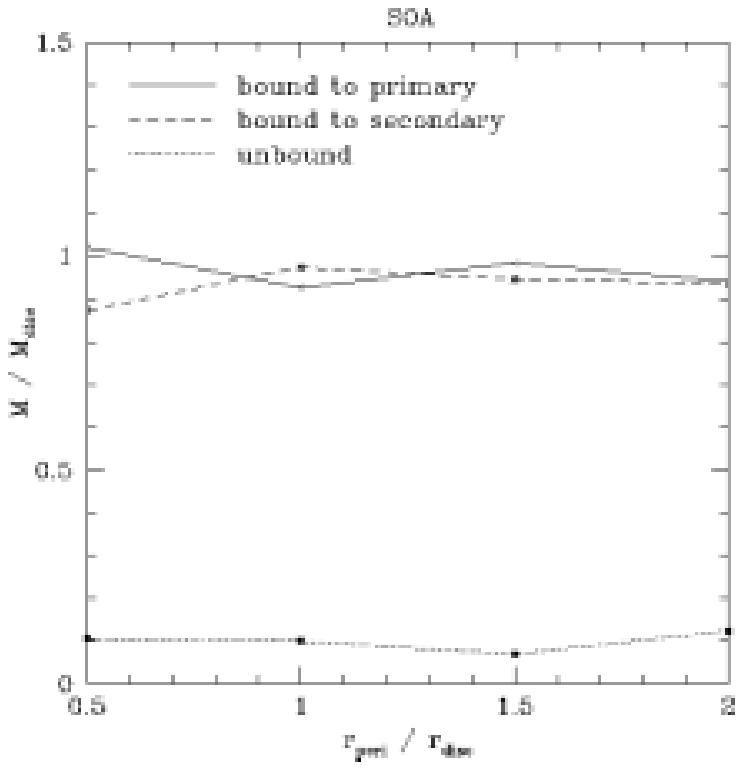}{\figpath/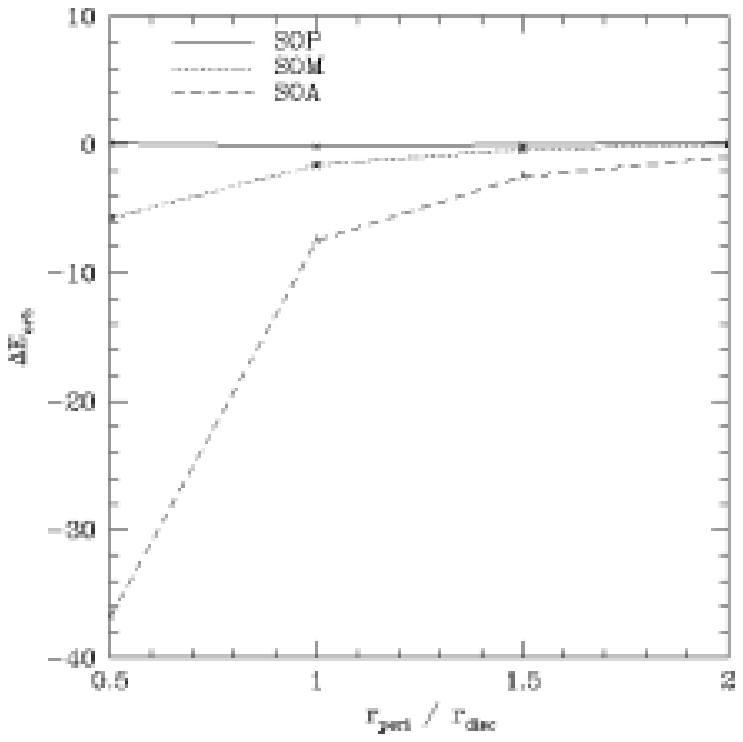}\\
\caption[Mass and energy transfer for coplanar disc-disc encounters]{\label{fig:cpdmass}Frames 1-3 show the fraction of the disc mass that ends up bound to the primary star, bound to the perturber and unbound, for SOP, SOM and SOA encounters respectively. Frame four shows the change in energy of the original orbit for all coplanar encounters.}
\end{figure*}

\section{Results}

A common mechanism operates during all of the coplanar disc-disc encounters. During the interaction, disc material is swept up into a shock that becomes gravitationally unstable and fragments. This fragmentation leads to the formation of new protostellar condensations. Some of these condensations are quickly disrupted by the original stars, others merge or are accreted directly onto the original stars; some, however, survive, and end up orbiting one of the original stars in a binary system, or are ejected from the system as single stars by a gravitational slingshot. Thus disc fragmentation caused by an interaction can lead to the formation of new stars and multiple systems.

At larger periastra, there is unsurprisingly a trend towards shorter shocks, and hence fewer fragments being formed, but with fragmentation and the formation of new stars still occurring. Even in the case of the grazing encounters, with $r_{peri}=2000$AU, the SOP and SOM interactions both end up with the formation of a binary companion to the primary star, and the SOA encounter forms companions to both stars.

The SOP encounters, in which the angular momenta of the discs and orbit are aligned, are possibly the most likely to occur in nature. For a binary system formed by the fragmentation of a single rotating core \cite{pringle89}, or by the fragmentation of a disc \cite{adams89}, such an alignment is expected. Turner et al.\ \shortcite{turner} have suggested that in star formation triggered by cloud-cloud collisions, the majority of encounters may be coplanar SOP encounters.

Frames 1-3 of Fig.~\ref{fig:cpdmass} show the masses that end up bound to each of the original stars, and the masses unbound from the system, for each of the coplanar encounters. The mass that is bound to a star may either be in the form of a circumstellar disc or bound up in newly-formed companions. For the SOP encounters, up to one-third of the total disc material present can become unbound from the system in the form of the tidally ejected spiral arms. Fragmentation occurs in the centre of the shock layer for SOP encounters, with seed noise determining which of the two original stars the fragments are captured by. The fragmentation produces only one or two condensations, and these  may then merge. The encounter can also trigger spiral instabilities within the discs, leading to more fragmentation at a later stage.

The SOP encounters tend to produce multiple systems, with the four encounters modelled producing a total of one single star, four binary systems and three triple systems. The triple systems are unlikely to survive for long, as they are not in an hierarchical configuration, and so each triple is likely to end up as a binary system and a single star that has been ejected from the group by the slingshot mechanism. The four SOP encounters are therefore likely to produce four single stars and seven binary systems, figures that are consistent with the observed binarity amongst pre-main-sequence stars -- although obviously many more simulations are needed to obtain a statistically robust estimate of this fraction.

Frame 4 of Fig.~\ref{fig:cpdmass} shows the change in the energy of the original orbit for the three types of coplanar encounters. Where fragmentation has occurred, each binary or triple that is formed is treated as being  a point mass at its centre of mass. The change in energy for the SOP encounters is small, although for encounters with $r_{peri} > r_{disc}$, the orbit becomes unbound, as was the case for the corresponding star-disc encounters.

Although two protostellar discs formed from the same core or disc might be expected to have their angular momenta aligned, this is not necessarily the case for encounters that occur when two protostars within the same cluster interact. In this case there is likely to be no preferred angle for the encounter. Observations by Hale (1994\nocite{hale}) indicate that for solar-type binaries with semi-major axes $a < 30\mbox{AU}$ there is a correlation between the directions of the orbit and the spins, but for binaries wider than this, and for hierarchical binaries, there is no correlation. In such cases,  SOP encounters are no more likely to occur than SOM or SOA encounters.

The SOM encounters produce longer shocks than the SOP encounters, and more protostellar condensations, but many of these are either disrupted or merge. The majority of the fragmentation occurs close to the primary star, and as a result more mass ends up bound to the primary than to the secondary, mostly in the form of companions. Less material is unbound from the system than for the SOP encounters. This can be seen in frame 2 of Fig.~\ref{fig:cpdmass}. The increased fragmentation also leads to the dissipation of more energy, particularly for encounters with $r_{peri} < r_{disc}$. The encounters with $r_{peri} / r_{disc}$ = 0.5 and 1.0 produce clusters of 5 and 9 stars respectively. 

The SOA encounters again produce long shocks. The shock fragments at either end, to produce one or two companions to each of the original stars. If at this stage a star acquires more than one companion, then these tend subsequently to merge to form a single companion, leaving the star in a close binary system. These close binary systems are bound to one another to form an hierarchical quadruple. Each close binary system orbits in the same sense as the original discs, while the binary systems orbit each other in the opposite sense, parallel to the original orbit. This occurs for all of the SOA encounters. For the closest encounter, with $r_{peri}=0.5r_{disc}$, the quadruple that forms is so close that the two new condensations are quickly tidally disrupted and then accreted by the original stars, leaving the end state of the system as a very tight binary.

The SOA encounters are the most dissipative of all the coplanar encounters. During the encounter, the angular momenta of the discs and the orbit, which are in the opposite sense, tend to cancel each other out, leaving the system tightly bound. A negligible amount of mass is unbound from the system.

As regards the energy change of the original orbits, the above results for disc-disc encounters are very similar to the results for star-disc encounters reported by Hall, Clarke \& Pringle\shortcite{hall}, and confirmed in Paper I:  retrograde encounters remove more orbital energy than prograde encounters. The SOP encounters, which are prograde for both discs, have very little effect energetically upon the orbit, and for $r_{peri}>r_{disc}$ can lead to an increase in the energy of the orbit. The SOM encounters, which are prograde for one disc and retrograde for the other, lead to the removal of similar amounts of energy from the orbit to the retrograde star-disc encounters (Fig.\ 6 in Paper I). The SOA encounters, in which the orbit is retrograde to both discs, causes the dissipation of much more orbital energy than either of the other types of encounter.

The disc-disc encounters presented here are markedly different from the corresponding disc-star encounters, in that they are dominated by the formation of a shock layer that fragments to produce multiple new companions to the original stars. However, the encounters can also lead subsequently to spiral-arm instabilities within the residual truncated discs, and these can cause fragmentation of the disc to produce additional companions -- in the same way that occured in the disc-star interactions. This secondary fragmentation occurs mainly in the medium-periastron disc-disc interactions ($1.0\leq r_{peri}/r_{disc}\leq 1.5$), since in the closest encounters the discs are completely destroyed, and in the furthest encounters the perturbation is generally not strong enough to lead to fragmentation. Fragmentation of the shock layer, on the other hand, occurs in all of the simulations, which have periastra out to $r_{peri}=2r_{disc}$. If the majority of young protostars undergo interactions while they still have massive discs, it is to be expected that these interactions will lead to formation of many lower-mass companions to the stars, so that the majority of the stars will end up in multiple systems.

\section{Conclusions}

Disc-disc encounters provide a mechanism for triggering disc fragmentation, and hence forming new protostars, as well as leading naturally to the presence of circumbinary discs. The results presented in this paper suggest that such encounters produce a binary fraction that is compatible with that observed amongst pre-main-sequence stars.

This mechanism is different from others proposed to explain the formation of multiple star systems. For instance, in the capture mechanism (Larson 1990\nocite{larson90}, Clarke \& Pringle 1991a, 199b, 1993\nocite{clarke:pringle91a}\nocite{clarke:pringle91b}\nocite{clarke:pringle93}), two protostars passing close to each other would dissipate orbital energy via a disc interaction and hence become bound. Here, the primary and secondary stars here may stay unbound from each other and end up in separate multiple systems. This mechanism is also different to the mechanism for binary formation by gravitational instability in a massive disc suggested by Adams et al.\ \shortcite{adams89}, in which the growth of an $m=1$ instability in an equilibrium disc eventually leads to fragmentation of the disc and formation of a single binary companion. Recent work \cite{laughlin:rozyczka} indicates that the operation of such instabilities would in fact tend to drive the disc towards stability and away from fragmentation. The mechanism identified here is  much more dynamic.
A key r\^{o}le is played by the formation and subsequent fragmentation of shock layers during the encounters. This shock fragmentation leads to the formation of new protostars and multiple star systems.

The circumbinary disc-like structures that are formed around the system in the SOP and SOM encounters are formed by the wrapping-around of spiral arms, rather than being the remains of an accreting envelope that is left around a binary \cite{artymowicz:lubow}. These discs are not in rotational equilibrium about the system, and so are likely to be transient features, although a stable circumbinary disc might result if there is an extended reservoir of material that can act to replenish the disc. 

It is expected that an average protostellar disc will undergo at least one disc-disc interaction during its lifetime. The results presented here show that such interactions lead to the formation of binary and multiple systems, as well as the presence of circumbinary material. Some of the stars retain truncated discs after the encounters, and have close companions with which they will undergo further disc-star and/or disc-disc interactions. This may in turn lead to the formation of new, closer companions. Hence interactions between massive extended protostellar discs can propagate binary star formation to smaller masses and shorter separations

In Paper III we investigate what happens when disc-disc interactions are non-coplanar, and carry out a full analysis of the results of simulations of disc-disc encounters.

\section*{Acknowledgements}

SJW and ASB acknowledge the receipt of University of Wales studentships. NF acknowledges the receipt of a PPARC studentship. HB and SJW acknowledge the support of PPARC post-doctoral grant GR/K94157. This work was supported by grants GR/K94140 and GR/L29996 from PPARC.

\end{document}